\documentclass[conference]{IEEEtran}

\usepackage{cite}
\usepackage[utf8]{inputenc}
\usepackage{graphicx}
\usepackage{float}
\usepackage{color, colortbl}
\usepackage{xcolor}
\usepackage{array}
\usepackage{multirow}
\usepackage{footnote}
\usepackage[normal]{threeparttable}
\usepackage{mathtools}
\DeclarePairedDelimiter{\ceil}{\lceil}{\rceil}

\usepackage[linesnumbered,vlined,norelsize,ruled,noend]{algorithm2e}

\SetCommentSty{mycommfont}

\makeatletter
\newcommand{\removelatexerror}{\let\@latex@error\@gobble}
\makeatother

\makeatletter
\def\blfootnote{\xdef\@thefnmark{}\@footnotetext}
\makeatother

\definecolor{color-eca+}{HTML}{00F9DE}
\definecolor{color-csma}{HTML}{FF9999}
\SetNlSty{bfseries}{\color{black}}{}

\makeatletter

\newcommand{\Rmnum}[1]{\expandafter\@slowromancap\romannumeral #1@}
\makeatother
\usepackage{graphicx,epsfig}
\usepackage{amsmath,amssymb,amsfonts}
\usepackage{mathrsfs}
\usepackage{algorithmic}
\usepackage{ulem}
\usepackage{authblk}
\usepackage{comment}
\newcommand*{\depaddr}[1]{\dagmark Computer and Information Sciences, \asmark Electrical and Computer Engineering} 

\newcommand*{\instaddr}[1]{\dagmark \asmark University of Delaware, \ddagmark Cisco Systems}

\newcommand*{\dagmark}[1][\dag]{\textsuperscript{\dag}}
\newcommand*{\asmark}[1][*]{\textsuperscript{*}}
\newcommand*{\ddagmark}[1][\ddag]{\textsuperscript{\ddag}}
\newcommand*{\email}[1]{\texttt{#1}}

\begin{document}


\title{Traffic Differentiation in Dense WLANs \\ with CSMA/ECA-DR MAC Protocol}

\author{%
Seyedmohammad Salehi\dagmark[1] ~~~ Li Li\asmark[1] ~~~ Chien-Chung Shen\dagmark[1] ~~~ Leonard Cimini\asmark[1] ~~~ John Graybeal\ddagmark[1] \\
\depaddr{\dagmark[1]}\\
\instaddr{\dagmark[1]}  \\
\email{\{salehi,lilee,cshen,cimini\}@udel.edu, jgraybea@cisco.com}
}

\maketitle

\begin{abstract}

In today's WLANs, scheduling of packet transmissions solely relies on the collision and success a station may experience.
To better support traffic differentiation in dense WLANs, in this paper, we propose a distributed reservation mechanism for the Carrier Sense Multiple Access Extended Collision Avoidance (CSMA/ECA) MAC protocol, termed CSMA/ECA-DR, based on which stations can collaboratively achieve higher network performance. In addition, proper Contention Window (CW) will be chosen based on the instantaneously estimated number of active contenders in the network. Simulation results from dense scenarios with traffic differentiation demonstrate that CSMA/ECA-DR can greatly improve the efficiency of WLANs for traffic differentiation even with large numbers of contenders. 
\end{abstract}


\section{Introduction}
\label{sec-intro}

To share wireless spectrum among contending Wi-Fi stations, Distributed Coordination Function (DCF) employs CSMA/CA with Binary Exponential Backoff. For traffic differentiation, Enhanced Distributed Channel Access (EDCA) specializes DCF parameters for different traffic Access Categories (ACs)  as well as giving more channel access time to high priority ACs by giving them different AIFS values and TXOP durations based on IEEE 802.11e.However, in dense scenarios, performance of EDCA severely degrades with increasing  number of contenders and resultant collisions \cite{sanabria2017high}. In particular, to satisfy QoS requirements, delay-sensitive traffic will blindly be given more channel access time, which may further increase the collision probability, and hence negatively affect the overall efficiency of the channel and QoS/QoE in dense scenarios \cite{sanabria2018traffic}.

Recently, there have been several efforts to improve CSMA/CA for emerging Wi-Fi standards \cite{sanabria2013future,da2016mac,sanabria2018traffic,sanabria2017high,syed2017performance,chun2012adaptive,xie2015adaptive,hong2012channel,khatua2016d2d}. Specifically, to address high collision rates in dense scenarios, different techniques have been developed to achieve collision-free schedules or reduce collisions by choosing optimal CW values. In particular, using CSMA Enhanced Collision Avoidance with Hysteresis and Fair Share (CSMA/ECA$_{\text{Hys+FS}}$) \cite{sanabria2017high,sanabria2018traffic,sanabria2013future}, upon a successful transmission, a station chooses the `expected value' of the Contention Window (CW) in which it just transmitted as its next backoff value (termed \textit{deterministic} backoff). This will gradually create a collision-free schedule in single traffic scenarios but is unable to achieve collision-free schedule in traffic differentiation.

Our goal is to further reduce collisions in both saturated and unsaturated dense scenarios with traffic differentiation. To achieve this goal, we extend CSMA/ECA$_{\text{Hys+FS}}$ \cite{sanabria2018traffic} to propose CSMA/ECA$_{\text{Hys+FS}}$ with {\em Distributed Reservation} (termed CSMA/ECA-DR$_{\text{Hys+FS}}$), where stations add their transmitting backoff stage (a 3-bit field) into transmitted MAC frame headers. Owing to the broadcast nature of wireless transmissions, other stations overhearing the transmitted frame(s) are also able to extract the current backoff stage of the transmitting station. Since in CSMA/ECA$_{\text{Hys+FS}}$, the current CW is derived from both backoff stage and $CW_{min}$, other stations can compute the future transmissions of the transmitting station (in terms of time slots) to avoid future `predicted' collisions with their own transmissions. CSMA/ECA-DR$_{\text{Hys+FS}}$ also employs Kalman filter \cite{bianchi2003kalman} to adjust the CW size of different traffic categories based on the estimated number of active contenders. Through extensive simulations, we show that CSMA/ECA-DR$_{\text{Hys+FS}}$ (hereafter, referred to as ECA-DR) outperforms CSMA/ECA$_{\text{Hys+FS}}$ (hereafter, referred to as ECA) in both saturated and non-saturated dense scenarios.

The rest of this paper is organized as follows. Related research is summarized in Section II. Section III describes ECA-DR in detail. Section IV presents simulation settings and traffic models used for the simulations followed by simulation results and evaluation. Section V concludes the paper with future research directions.

\section{Related Work}
\label{sec-related}

The closest work to ECA-DR is EBA \cite{choi2005eba} which is designed to improve CSMA/CA with a single type of traffic. Using EBA, a station announces to other stations its future random backoff value and current offset (so that recipients can synchronize their reservation windows) in a 24-bit field piggy-backed on each transmitted MAC frame header. A receiving station should keep a reservation window to compute its next backoff value based on the announcement overheard and choose the offset if it is not an active transmitter (i.e., no packet\footnote{Packet and frame are used interchangeably in this paper and both refer to MAC layer PDU (M-PDU).} in its queue). However, the backoff value is chosen randomly and, hence, would not help to reduce collisions among a large number of contenders. 

Instead, a station using ECA-DR shares its {\em backoff stage} value with other stations, in a {\bf 3-bit} field added to the MAC frame header. Since ECA-DR is based on ECA, a station chooses the expected value of CW in which it transmitted its frames and hence, other stations overhearing the transmitting station could compute {\em both} the future transmission time of the transmitting station {\em and} the number of frames to be transmitted. Instead of keeping a reservation window as in EBA (which incurs higher memory usage and computational cost), each station, using ECA-DR, only keeps the prohibited backoff values, decreases them with its own backoff values (if it has packets to transmit), and avoids choosing them for its future transmissions. 

In addition, adaptive selection of optimal CW size based on different measured criteria in the network has been shown to improve the performance of CSMA/CA. Such criteria can be the
number of contenders estimated based on Conditional Collision Probability ($\rm{P_{cc}}$) \cite{bianchi2003kalman,syed2017performance}; 
the number of contenders and average idle slots \cite{chun2012adaptive}; the
channel utilization ratio and retransmit counts \cite{xie2015adaptive}; the
channel Bit Error Ratio (BER), backoff parameters and contention level \cite{deng2008contention,hong2012channel}; 
the delay deviation and channel congestion status \cite{khatua2016d2d}, etc. 
Specifically, ECA-DR uses $\rm{P_{cc}}$ to approximate the number of active contenders and selects the appropriate backoff stage for different traffic categories both between two unsucessful transmissions and when the backoff stage is supposed to be reset to zero.

\section{Description of ECA-DR}
\label{sec-eca-DR}
	
Algorithm \ref{alg:ECA-DR}\footnote{This algorithm is borrowed from \cite{sanabria2017high} and adapted to traffic differentiation using ECA-DR.} without the blue lines represents ECA with traffic differentiation. It is important to note that usage of AIFS violates the assumption that all backlogged stations simultaneously decrease their backoff values after each slot, and hence, AIFS is not practical in ECA and ECA-DR \cite{barcelo2009traffic}. Unlike CSMA/CA in which successive transmissions of a station have no correlation with each other (even after successful transmissions, the backoff stage ($\rm{k_i}$) will be reset to zero), ECA would relate successive successful transmissions of a station by the ``Hysteresis'' mechanism (i.e., choosing the expected value of the transmitting CW as the next backoff value, line \ref{hysteresis}).
The issue of fairness among stations that wait longer to transmit is also addressed by the ``Fair Share'' mechanism which relates the number of transmitting frames to the transmitting backoff stage, $\rm{k_i}$ (line \ref{fairshare}).

ECA only reaches a collision-free schedule in saturated single traffic scenarios, since almost all contenders will converge to the same CW. However, ECA is unable to converge to a collision-free schedule in traffic differentiation or unsaturated scenarios within dense deployments. In traffic differentiation, delay-sensitive access categories (AC) require frequent transmissions and hence shorter transmission intervals (and CW sizes). However, among the deterministic backoff values computed by Algorithm \ref{alg:ECA-DR} (i.e., 3, 7, 15, 31, 63, 127, 255, 511), 15 is divisible by 3 (AC[VO], the voice AC, in backoff stage 0) and 63 is divisible by 7 (AC[VO] in backoff stage 1 and AC[VI], the video AC, in backoff stage 0). Thus, AC[VO] and AC[VI] might collide with other traffic categories or the same traffic categories with larger deterministic backoff values.In unsaturated scenarios, due to frequent queue flushes, a station might not retain its deterministic backoff value for a long time. This is followed by choosing random backoff values with resultant collision increase in dense scenarios. The proposed distributed reservation mechanism of ECA-DR correctly identifies these collisions and prevents them from happening. 

\begin{algorithm}[tb]
 \While{the device is on}
 {
  $\rm{R \leftarrow 6; K \leftarrow 5; AC \leftarrow 4}$\; 
  $\rm{BIV = 1}$;  \qquad \qquad \qquad \quad \tcp{BOS increase value}
  $\rm{{CW}_{min}[AC] \leftarrow [32,32,16,8]}$\;
  \For{$\rm{i} \gets 0$ \KwTo $\rm{AC}-1$}{
   $\rm{r_i \leftarrow 0; \rm{k_i} \leftarrow 0}$\; 
   $\rm{B_i} \leftarrow \mathcal{U}[0,2^{k_i}\rm{CW}_{min}[i]-1]$\;
  }
   \While{there is a packet in $\rm{Q_i}$ to transmit}{
    \Repeat{($\rm{r_i = R}$) or (success)}{
     \While{$\rm{B_i>0}$}{
      wait 1 slot\;
      $\rm{B_i \leftarrow B_i-1}$\;
      \textcolor{blue}{
       \If {overheard a packet \rm{(p)}}{
        $\rm{NT \leftarrow (2^{p.b}\rm{CW}_{min}[p.AC])/2-1}$\;
        \If {$\rm{B_i = NT}$}{
         $\rm{B_i \leftarrow \mathcal{U}[0,2^{k_i}\rm{CW}_{min}[i]-1]}$\; 
        }
       }
      }
     }
     Transmit $\rm{2^{k_i}}$ packet(s); \label{fairshare} \qquad \: \tcp{Fair Share} 
     \If{collision}{
      $\rm{r_i \leftarrow r_i+1}$\;
      \textcolor{blue}{
       Choose $\rm{BIV}$ based on Estimated NAC\;
       }
      $\rm{k_i \leftarrow \min (k_i+BIV, K)}$\;
      $\rm{B_i \leftarrow \mathcal{U}[0, 2^{k_i}  \rm{CW}_{min}[i] -1]}$\;
     }
    }
	$\rm{r_i \leftarrow 0}$\;
	\eIf{success}{
	 $\rm{B_i \leftarrow (2^{k_i}\rm{CW}_{min}[i])/2-1}$; \label{hysteresis} \hfill \tcp{Hysteresis}  
	}
	{
	Discard $\rm{2^{k_i}}$ packet(s)\;
	$\rm{k_i \leftarrow 0}$\;
    \textcolor{blue}{
     Choose $\rm{k_i}$ based on Estimated NAC\;
	}
    $\rm{B_i \leftarrow \mathcal{U}[0, 2^{k_i} \rm{CW}_{min}[i]-1]}$\;
   }
  }
 Wait for a packet in $\rm{Q_i}$ to transmit\;
 $\rm{k_i \leftarrow 0}$\;
 \textcolor{blue}{
  Choose $\rm{k_i}$ based on Estimated NAC\;
 }
 $\rm{B_i \leftarrow \mathcal{U}[0, 2^{k_i} \rm{CW}_{min}[i]-1]}$\;
}
\vspace{0.2cm}
\caption{ECA-DR with Traffic Differentiation}
\label{alg:ECA-DR}
\end{algorithm}

\subsection{Distributed Channel Reservation}
The blue-colored section in Algorithm \ref{alg:ECA-DR} depicts the general inner workings of ECA-DR and its integration with ECA. 
To compute the next transmission time (in terms of time slots) of a transmitting station, an overhearing station requires the knowledge of the type of overheard traffic, its CW$_\text{min}$ and the transmitting backoff stage. The type of traffic can be extracted from the ``TID'' subfield of the ``QoS Control'' field in the MAC header (denoted by p.AC in line 14) and CW$_\text{min}$ for that type of traffic is based on IEEE 802.11e. Also, the added 3-bit field to the MAC frame header contains the transmitting backoff stage (p.b). Line 14 shows the computation of the next transmission time of an overheard frame. After computing the next transmission time, the overhearing station compares the next transmission time with its own backoff values of backlogged traffic categories (line 15).
If a collision is predicted with any of the station's traffic categories, the station should choose another random backoff value.

The 3-bit field can represent numbers 0 to 7 in binary, but the maximum backoff stage will not exceed 5 or 6 (5 in our simulations). If a station finds its queue empty, it will announce its queue status (similar to setting the "More Data" subfield of the "Frame Control" field) by setting the 3-bit field to 7 (i.e., 111 in binary) in order not to prevent other stations from choosing the station's deterministic backoff value computed in Algorithm \ref{alg:ECA-DR} as NT. If the backoff stage value is less than 7, the next transmission (NT) will be added to the list of prohibited backoff values kept by each station. If a station requires a random backoff value for any of its traffic categories (new packet in an empty queue or after collision), it avoids choosing the prohibited values. Note that once a prohibited value is added to the list of prohibited values of a station, a station should count it down with its own backoff values for each passing time slot. If the station does not have any packet in its queues, it should still count down the prohibited values for each passing time slot. 

In our simulations we put this 3-bit field in the Address 4 field of MAC frame header so as to avoid any MAC frame header overhead. Including transmitting backoff stage into the MAC frame header also plays the role of RTS control frame. In order to avoid hidden terminals in multi-hop and overlapping basic service set (OBSS) scenarios, a receiver should also include this field in the ACK control frame to also play the role of CTS control frame. Unlike RTS/CTS that reserves the channel for a transmission that follows CTS, the distributed reservation mechanism of ECA-DR only instructs the overhearing stations to refrain from transmission at the next transmission of a transmitting station (i.e., the overhearing stations may transmit before or after). Therefore, distributed reservation does play the role of RTS/CTS without four-way handshake which is important for short frames and delay-sensitive applications.
\subsection{Backoff Stage Selection}
Conditional Collision Probability (P$_\text{cc}$) is the probability of occurring packet collisions that are only caused by transmissions.
Based on \cite{bianchi2003kalman}, P$_\text{cc}$ will remain constant irrespective of the number of packet retransmissions by a station and can be used to infer the number of active contenders (NAC) in the network. P$_\text{cc}$ can be computed by dividing the number of busy and collision slots (a station overhears) by the total slots:
\begin{equation}
\rm{P_{cc}} = \frac{\rm{Slot_{busy}+Slot_{Collision}}}{\rm{Total Slots}}
\end{equation}

Also we noticed in dense networks, always resetting backoff stage to zero (after receiving packet in an empty queue or packet drop after reaching the maximum retry limit) highly contribute to the overall collisions and performance degredation. Thus, a station can compute P$_\text{cc}$ to choose the proper CW when it is supposed to reset its backoff stage. In Algorithm \ref{alg:ECA-DR}, ECA-DR uses P$_\text{cc}$ to estimate the NAC and choose the proper CW (1) between two consecutive transmissions of a station if the first transmission results in a collision by computing the backoff stage increae value (lines 20) and (2) where the station has to reset its backoff stage (line 30 and line 34). 

Optimal CW selection cannot be used with the ``hysteresis'' mechanism of ECA because deterministic backoff values chosen after successful transmissions might divide each other which may cause more collisions in dense networks. Thus, to compute `proper' CW for AC[BE] and AC[BK], ECA-DR chooses the backoff stage (k$_{AC}$ and BIV) based on Eq. 2:
\begin{equation}
\rm{2^{k_{AC}}CW_{min}[AC] > NAC^2 \cdot P_{cc}}
\end{equation}
Due to stringent delay requirements of delay-sensitive traffics, CW for AC[VO] and AC[VI] is chosen equal to half of the value computed by Eq. 2.

\section{Simulation and Analysis}
\label{sec-evaluation}

\begin{figure*}[tb]
	\centering
		\includegraphics[width=18.3cm]{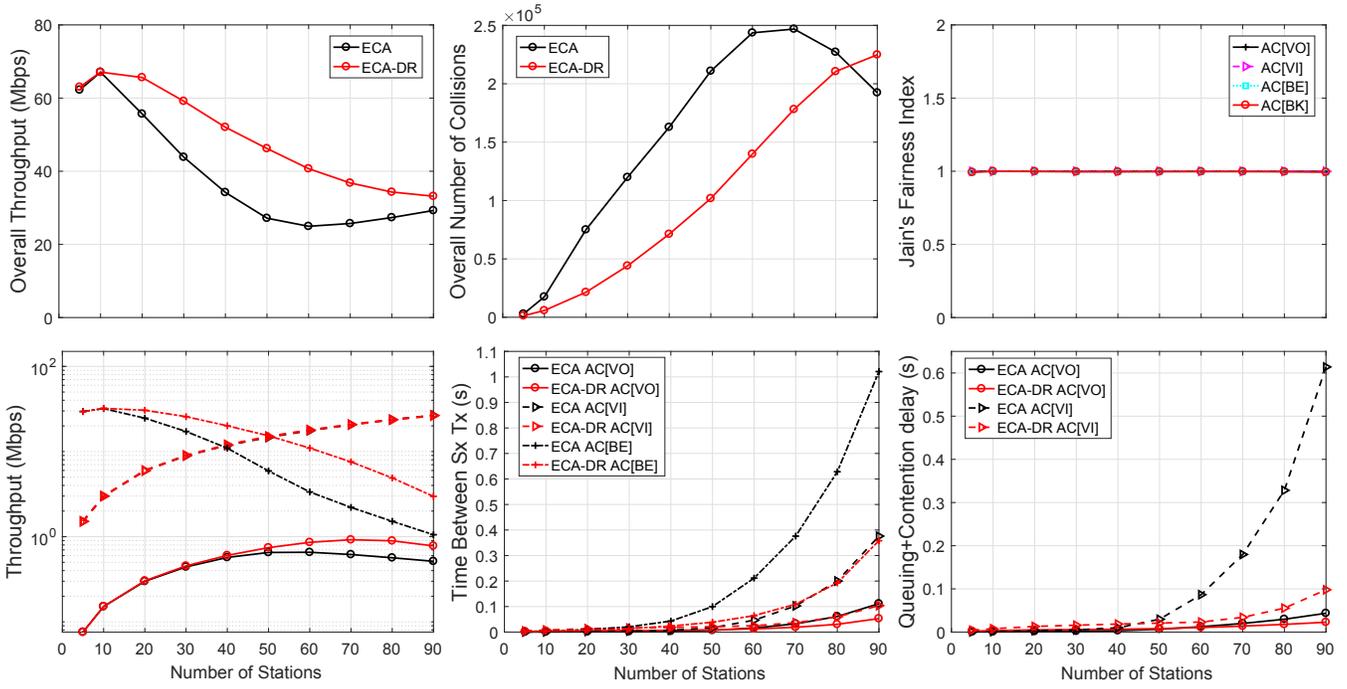}
		\caption{\label{fig:bebk-sat} Results of saturated scenario for 5-90 stations each with 4 ACs. Overall network throughput, collisions, and Jain's fairness index are shown on top. Results for different ACs are shown on the bottom. Legend of throughput per AC can be inferred from the next sub-figure.}

\end{figure*}


\begin{table}
		\centering
		\caption{Other parameters for the simulations}
		\label{tab:mac-params}
		\begin{tabular}{cc}
			\hline
			{\bfseries Parameters} & {\bfseries Value}\\
			\hline
			Physical channel rate & 65~Mbps\\
            \hline
            Channel width & 20 MHz \\
            \hline
            Number of streams & 2x2 MIMO \\
			\hline
            Empty slot duration & $9~\mu s$\\
			\hline
            DIFS & $28~\mu s$\\
			\hline
            SIFS & $10~\mu s$\\
			\hline
            Maximum retransmission attempts & 6\\
			\hline
            Packet size & 1470~Bytes\\
			\hline
            MAC queue size & 2000~Packets\\
			\hline
		\end{tabular}
	\end{table}


\begin{figure*}[tb]
	\centering
		\includegraphics[width=18.3cm]{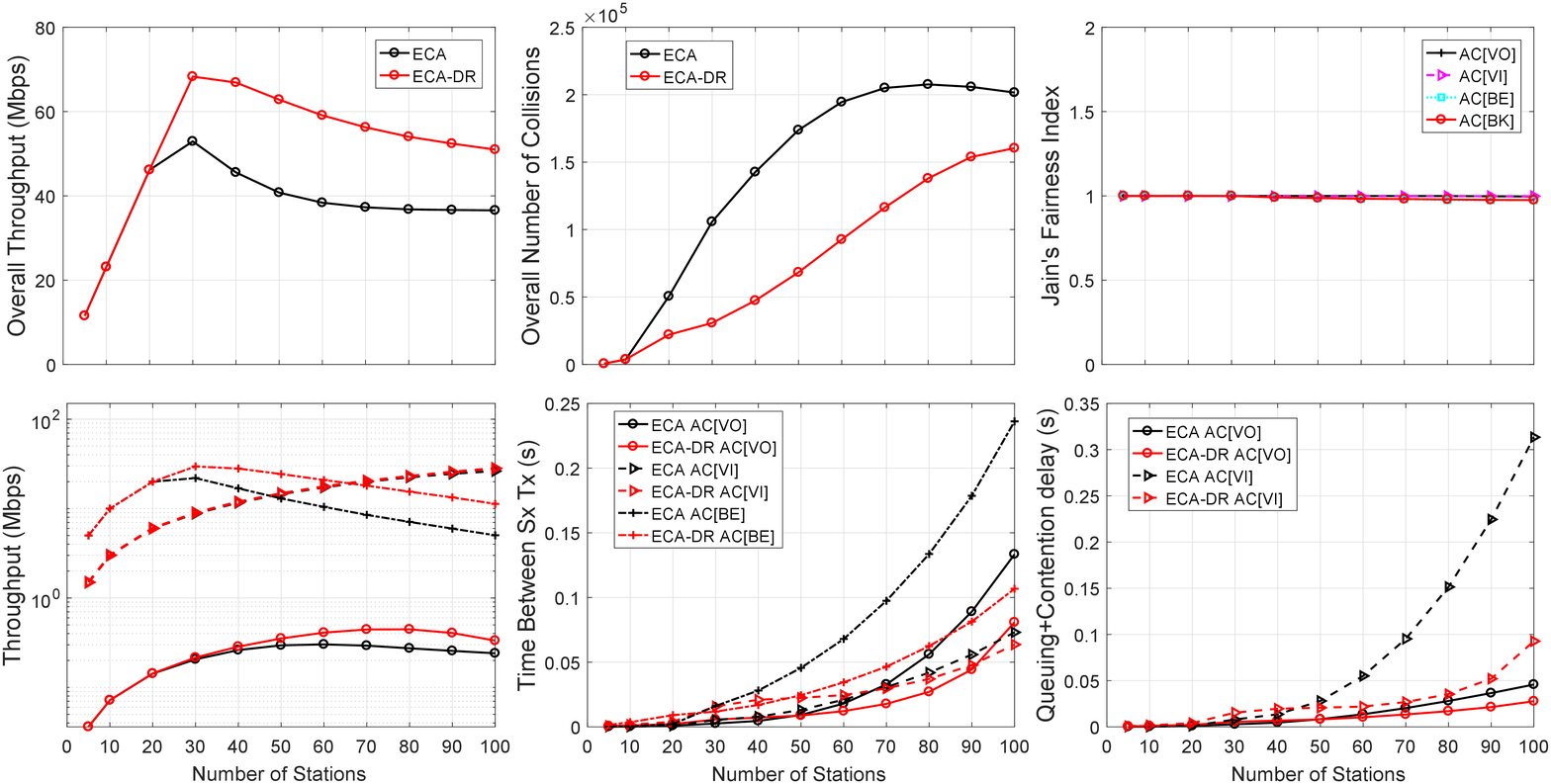}
		\caption{\label{fig:unsat} Results of unsaturated scenario for 5-100 stations each with 4 ACs. Overall network throughput, collisions, and Jain's fairness index are shown on top. Results for different ACs are shown on the bottom. Legend of throughput per AC can be inferred from the next sub-figure.}
\end{figure*}


We extend \cite{sanabria2018traffic} to implement ECA-DR in the COST simulator \cite{chen2002reusing}. Simulations are carried out in a single hop scenario where all stations are in transmission range of each other and the channel is assumed to have no errors. 

\subsection{Simulation settings}

To resemble dense traffic scenarios, each station is simulated to have four types of traffic (i.e., AC[VO], AC[VI], AC[BE] and AC[BK]). AC[VI] source traffic is based on H.264/Advanced Video Coding (H.264/AVC) with compression mechanism and resulting rate-variability. AC[VO] is chosen based on Internet Low Bit Rate Codec (iLBC) with silent detection (payload of 38~bytes with 20~ms intervals). For more information about the detail of Voice and Video codecs, we refer the readers to \cite{sanabria2018traffic}. In \textit{saturated} settings, MAC queues of AC[BE] and AC[BK] always have packets to transmit with packet arrival rate of 65~Mbps which is larger than the throughput they can attain. In \textit{unsaturated} traffic scenarios, the packet arrival rate to the MAC queues of AC[BE] and AC[BK] is 1~Mbps that will result in frequent queue flushes. 

Simulations are based on 10 repetitions with different seeds that simulate 60 seconds of different protocols (i.e., ECA and ECA-DR). Other simulation parameters are illustrated in Table \ref{tab:mac-params}. Since the goal is to use the proposed distributed reservation mechanism instead of the expensive RTS/CTS mechanism in the conventional ways, all simulations are carried out without RTS/CTS. Thus, in saturation scenarios, the transmission duration of a successful frame can be computed by:
\begin{equation} \label{tsuccess}
\rm{T_{success}} = \rm{T_{frame}} + SIFS + \rm{T_{BlockACK}} + DIFS + \rm{T_{\sigma}},
\end{equation}
where T$_{\sigma}$ is the duration of an empty slot. T$_{\text{frame}}$ and T$_{\text{BlockACK}}$ are computed from Eqs. \ref{tframe} and \ref{tba}, respectively, as follows.
\begin{equation} \label{tframe}
\rm{T_{frame}} = \rm{T_{PHY}} + \ceil[\Bigg]{\frac{SF+k(MD + L_{MH} + L_{data})+TB}{OFDM~Rate}} T_{sym}
\end{equation} 
In Eq. \ref{tframe} T$_\text{PHY}$ is 32~$\mu$s, Service Field (SF) is 2~bytes, k is the transmitting backoff stage that gives the number of aggregated MAC frames (A-MPDUs), MPDU Delimiter (MD) is 4~bytes, length of MAC header (L$_\text{MH}$) is 36~bytes including 3-bit field, Tail Bits (TB) is 6~bits and the duration of OFDM symbol T$_\text{sym}$ is 4~$\mu$s. OFDM Rate is computed based on the number of  subcarriers (234 for 20~Mhz bandwidth), the number of bits per OFDM symbol (6), coding rate (3/4) and antenna settings (MIMO). 
\begin{equation} \label{tba}
\rm{T_{BlockACK}} = \rm{T_{PHY}} + \ceil[\Bigg]{\frac{SF+L_{BlockACK}+TB}{OFDM~Rate}} T_{sym}
\end{equation}
Length of Block Acknowledgement (L$_\text{BlockACK}$) is 32~bytes.

\subsection{Analysis of simulation results}
Fig. \ref{fig:bebk-sat} depicts the results of ECA-DR in saturated scenarios. 

The top 3 sub-figures show the overall statistics. The overall number of collisions in ECA-DR is less than half of that in ECA for up to 50 users. Not only does ECA-DR achieve higher throughput than ECA (top-left sub-figure), it also reduces the average delay of delay-sensitive traffic (bottom-right sub-figure). This will give the chance to support larger numbers of Voice (AV[VO]) and real-time Video AC[VI] users. Average queuing delay and time between successful transmissions of AC[VO] remain below 10 ms for up to 70 users and for  AC[VI], remain below 100 ms for up to 90 users. ECA can only satisfy delay requirements of 60 AC[VO] users and 60 AC[VI] users.

Real-world scenarios are mostly represented by \textit{non-saturated} scenarios. AC[BE] and AC[BK] represent web-surfing, email and file download (not all the users constantly saturate these ACs). Packet arrival rate to the MAC queue of these traffics is 1~Mbps. Fig. \ref{fig:unsat} illustrates the network statistics based on different network metrics. Like saturated scenario, ECA-DR outperforms ECA in achieving both higher throughputs and lower delays. In unsaturated scenarios, ECA-DR satisfies delay requirements of 100 AC[VI] and 60 AC[VO] users. ECA can only satisfy delay requirements of 70 AC[VI] and around 55 AC[VO] users.

\section{Conclusions and future directions}
\label{sec-conclusion}

Three features of CSMA/CA that cause network performance degradation in dense networks are (1) prioritization of delay sensitive traffics by blindly giving them more channel access time, (2) doubling of CW after collision to reduce further collisions without considering the number of active contenders, and (3) scheduling of packet transmissions by solely relying on the collision or success a station may experience. 
Choosing optimal CW based on instantaneous estimated number of active contenders has been shown to improve the efficiency of CSMA/CA-based MAC protocols. However, optimal CW selection cannot be used with the ``hysteresis'' mechanism of ECA due to asynchronous stations decisions.
In this paper, we introduced a distributed reservation mechanism for ECA, termed ECA-DR, based on which stations can collaboratively achieve higher network performance. To be able to use ECA with proper CW, ECA-DR chooses the CW based on estimated number of active contenders. Simulation results demonstrate that ECA-DR can greatly improve the efficiency of WLANs and, hence, support larger numbers of voice and real-time video users.

We plan to reduce inter-cell interference in the overlapping basic service set (OBSS) using ECA-DR and also to extend ECA-DR to multi-hop scenarios and study the impact of hidden terminals without the explicit use of RTS/CTS.

\bibliographystyle{IEEEtran}

\end{document}